\documentstyle[12pt]{article}
\begin{document}
\catcode`@=11
\def\seceqaa{\@addtoreset{equation}{section}
\def\theequation{A\arabic{equation}}}
\catcode`@=12
\title{Relating Green's Functions in Axial and Lorentz 
Gauges using Finite Field-Dependent BRS Transformations}
\author{{Satish. D. Joglekar}\thanks{e-mail:sdj@iitk.ac.in},
{A. Misra} \thanks{e-mail:aalok@iitk.ac.in}\\
Department of Physics, Indian Institute of Technology, \\
Kanpur 208 016, UP, India}
\maketitle
\vskip 0.5 true in 
\begin{abstract}
We use   finite field-dependent BRS transformations (FFBRS) to connect
the Green functions in a set of two otherwise unrelated  gauge choices.
We choose the Lorentz and the axial gauges as examples. We show how the
Green functions in axial gauge can be written as a series in terms of
those in Lorentz gauges. Our method also applies to operator Green's functions.
We show that this process involves another set of related
FFBRS transfomations that is derivable from 
infinitesimal FFBRS. We suggest possible applications.

\end{abstract}


\section{Introduction}

Strong, weak and electromagnetic interactions are known to be described
very well by the standard model [SM] \cite{cl} which is a nonabelian gauge
theory. Calculations in nonabelaian gauge theories require a choice of 
gauge. These can be chosen in many ways. There are many families of
gauges that have been used in practical calculations. Lorentz-type gauges
\cite{cl} have been used in a large number of calculations in SM on 
account of their covariance and availability of a free gauge
parameter that helps in the check of gauge-independence. Another
family of gauges, the axial gauges, $\eta\cdot A=0$ have also
been used extensively \cite{lm}. These have the formal
audvantage of being free of ghosts, which leads to simplifications
in calculations of Green's functions, anomalous dimensions, etc. Special
cases such as those with $\eta^2=0$, viz the light cone gauges have been
used extensively   in perturbative QCD calculations \cite{l}. Analogous
gauges, the planar gauges have the advantages of the axial gauges and
in addition  have simpler propagator and hence have also found
favor \cite{s}. The radial gauges have found widespread use in the 
context of QCD sum rules and operator product expansions in QCD
\cite{dn}. Certain quadratic gauges have been found to simplify
Feynman rules and  calculations of diagrams in spontaneously broken gauge
theories [SBGT] \cite{cl}. $R_\xi$ gauge have extensively been
used in performing practical calculations and in formal  
arguments in SBGT \cite{al}. Thus, to summarize,
various descriptions of gauge theories have been found useful in various
different contexts. It therefore becomes an 
important question how the calculations in various (families of)
gauge choices are related to each other.

Now, we expect the physical results to be independent of the choice of 
gauge. Indeed, gauge-independence in a limited framework,
has been proven in early days \cite{al}, \cite{h}. For example,
within the Lorentz-type of gauges, one establishes
the $\lambda$-independence of the physical
observable, etc. \cite{al}. Such proofs utililize
the {\it infinitesimal} gauge transformations reponsible
for gauge-parameter change. Ways of connecting Green's
functions in a family of gauges (and establishing explicitly gauge-independence)
has not been done until recently. Indeed, recently discrepancies have
been reported in anomalous dimension calculations in the Lorentz-type and
axial-type gauges \cite{jb1}.

Thus, for these and further reasons detalied below, we consider
it valuable to obtain a procedure to connect the Green's functions 
in different families of gauges. Certain progress along these
lines has already been made. In \cite{jm}, we established a
general procedure for obtaining a field transformation that 
connects the vacuum-to-vacuum amplitude $W$ (and also the
vacuum-expectation values of  gauge invariant observables) in two sets of
gauges. This was elaborated by a set of  examples given there.
These transformations turned out to be a generalization of the usual
BRS transformations in which the anticommuting global parameter
is (i)field-dependent, but $x$-independent, and (ii) finite rather than
infinitesimal. These were thus named finite 
field-dependent BRS (FFBRS) transformations. In view of the importance
of the two families of gauges, viz the Lorentz-type
and the axial-type practical in calculations, 
we established a similar FFBRS connections between
these set of gauges \cite{jb}, \cite{talk}.

In this work, we establish a connection between
arbitrary Green's functions (or operator Green's
functions) in two sets of gauges, and in view of their practical 
importance, we choose these to be the
Lorentz and the axial-type gauges. Of course, once an
FFBRS is established between any two sets of gauges, an identical
procedure would go through.  We show that the required procedure
involves another FFBRS.  We establish  finally
a compact result expressing an arbitrary Green's function/
operator Green's function in axial  gauges with a closed expression
involving similar Green's functions in Lorentz
gauges. The expression can then be evaluated in principle, as a power
seris in $g$ to desired order.

We shall mention in passing a number of applications of this result. 
We can use the result  for the axial gauge propagator in terms of
the Lorentz gauge propagator as a way for obaining the
prescription for the ${1\over{\eta\cdot k}}$-singularity.
This is so since we understand how to deal
with the Green's
functions in Lorentz-type gauges. We should also
be able to eliminate the possible reported discreprancy between
the anomalous dimensions of physical observables \cite{al}
in the two sets of gauges.  Such and other possible applications are
under progress.

We now summarize the plan of the paper. In Section 2,
we review the background needed together with the results of 
references \cite{jm}-\cite{talk} on FFBRS transformations. In Section 3, we
show how Green's functions in the two sets of gauges can be related. 
We show how this involves the
use of another FFBRS. Appendix A deals with FFBRS along
the lines of \cite{jm}. Section 3 gives a compact formula
relating the Green's functions in the two gauges. 
Section 4 gives a simple example of the compact
formula obtained in Section 3.
Section 5 deals with some future intended applications and conclusions.

\section{Summary of Results on FFBRS Transformation between Lorentz
and Axial-type Gauges}
\subsection{Notations and Conventions}

We start with the Faddeev-Popov effective action (FPEA) in linear 
Lorentz-type gauges:
\begin{equation}
\label{eq:FPEAL}
S^L_{\rm eff}[A, c, {\bar c}]=\int d^4x\biggl(-{1\over 4} F^\alpha_{\mu\nu}
F^{\alpha, \mu\nu}\biggr)+S_{\rm gf}+S_{\rm gh},
\end{equation}
where the gauge-fixing
action $S_{\rm gf}$ is given by:
\begin{equation}
\label{eq:gfL}
S^L_{\rm gf}=-{1\over{2\lambda}}
\int d^4x\sum_{\alpha}(\partial\cdot A^\alpha)^2\equiv-{1\over{2\lambda}}
\int d^4x\sum_{\alpha}(f^\alpha_L[A])^2,
\end{equation}
and the ghost action $S_{\rm gh}$ is given by:
\begin{equation}
\label{eq:ghL}
S^L_{\rm gh}=-\int d^4x {\bar c}^\alpha M^{\alpha\beta}c^\beta,
\end{equation}
where
\begin{equation}
\label{eq:MLdef}
M^{\alpha\beta}[A(x)] \equiv \partial^\mu D^{\alpha\beta}_\mu(A,x).
\end{equation}
The covariant derivative is defined by:
\begin{equation}
\label{eq:covDdef}
{\rm D}^{\alpha\beta}_\mu\equiv\delta^{\alpha\beta}\partial_\mu
+gf^{\alpha\beta\gamma}A_\mu^\gamma.
\end{equation}

In a similar manner, the FPEA in axial-type gauges, is given by:
\begin{equation}
\label{eq:FPEAA}
S^A_{\rm gf}\equiv-{1\over{2\lambda}}\int d^4x\sum_{\alpha}
(\eta\cdot A^\alpha)^2\equiv-{1\over{2\lambda}}\sum_{\alpha}\int d^4x 
(f^\alpha[A])^2.
\end{equation}
We require $\eta_\mu$ to be real, but otherwise unnrestricted.
and
\begin{equation}
\label{eq:ghA}
S^A_{\rm gh}=-\int d^4x {\bar c}^\alpha\tilde{M}^{\alpha\beta}c^\beta,
\end{equation}
with
\begin{equation}
\label{eq:tildeMdef}
\tilde{M}^{\alpha\beta}=\eta^\mu D^{\alpha\beta}_\mu.
\end{equation}
In the $\lambda\rightarrow 0$ limit, 
\begin{equation}
\label{eq:lambdazeroom}
e^{iS^A_{\rm gf}}\sim \prod_{\alpha,x}\delta
\biggl(\eta\cdot A^\alpha(x)\biggr).
\end{equation}
Thus, in the presence of the delta function, the $A$-dependent term in 
$\tilde{M}$ can be dropped leading to the formally ghost-free matrix.
As is well known, $S^L_{\rm eff}$ and $S^A_{\rm eff}$ 
are invariant under the BRS transformations:
\begin{eqnarray}
\label{eq:BRS1}
& & \delta
 A^\alpha_\mu(x)=D_\mu^{\alpha\beta}c^\beta(x)\delta\Lambda
\nonumber\\
& & \delta c^\alpha(x)=-{g\over 2}f^{\alpha\beta\gamma}c^\beta(x)
c^\gamma(x)\delta \Lambda
\nonumber\\
& & \delta{\bar c}^\alpha(x)={{f^\alpha[A]}\over{\lambda}}\delta\Lambda,
\end{eqnarray}
where $f^\alpha[A]=\partial\cdot A^\alpha$ or $\eta\cdot A^\alpha$,
depending  on whether one  has written the action in the Lorentz or the axial-type gauges.

\subsection{FFBRS Transformations}

As observed by Joglekar and Mandal \cite {jm}, in (\ref{eq:BRS1}),
$\delta\Lambda$ need not be infinitesimal
 nor need it be field-independent as long as it  does not  depend on
$x$ explicitly for (\ref{eq:BRS1}) to be a symmetry of FPEA
In fact, the following finite field-dependent BRS (FFBRS)
transformations were introduced:
\begin{eqnarray}
\label{eq:BRS2}
& & 
{{A^\prime}}\ ^\alpha_\mu = A^\alpha_\mu  
+ D_\mu^{\alpha\beta}c^\beta(x)\Theta[\phi]
\nonumber\\
& & {{c^\prime}}\ ^\alpha = c^\alpha-{g\over 2}f^{\alpha\beta\gamma}c^\beta(x)
c^\gamma(x)\Theta[\phi]
\nonumber\\
& & {{\bar c^\prime}}\ ^\alpha = {\bar c}^\alpha 
+ {{f^\alpha[A]}\over{\lambda}}\Theta[\phi],
\end{eqnarray}
or generically
\begin{equation}
\label{eq:BRS3}
\phi^\prime_i(x)=\phi_i(x)+\delta_{\rm BRS}\phi_i(x)\Theta[\phi],
\end{equation}
where $\Theta[\phi]$ is an $x$-independent functional of $A,\ c,\ {\bar c}$ (generically denoted
by $\phi_i$) and these were also the symmetry of the FPEA.  
The transformations of the  form ({\ref{eq:BRS2}) were used to
connect actions of different kinds for Yang-Mills theory in \cite {jm} and 
\cite {jb}. The FPEA is invariant under (\ref{eq:BRS2}), but the
functional measure is not invariant under the
(nonlocal) transformations (\ref{eq:BRS2}). The Jacobian for the
FFBRS transformations can be expressed (in special
cases dealt with in \cite{jm,jb}) effectively as
$\exp(iS_1)$ and this $S_1$ explains the difference between the two
effective actions. Such FFBRS transformations were constructed in 
\cite{jm},\cite{jb} by
integration of an infinitesimal field-dependent BRS
(IFBRS) transformation:
\begin{equation}
\label{eq:BRS4}
{d\phi_i(x,\kappa)\over{d\kappa}}
=\delta_{\rm BRS}[\phi(x,\kappa)]\Theta^\prime[\phi(x,\kappa)]
\end{equation}
The integration of (\ref{eq:BRS4}) from $\kappa=0$ to 1, leads to the FFBRS transformation
of (\ref{eq:BRS3}) 
with $\phi(\kappa=1)\equiv \phi^\prime$ and $\phi(\kappa=0)=\phi$.
Further $\Theta$ in (\ref{eq:BRS3}) was related to $\Theta^\prime$ by:
\begin{equation}
\label{eq:BRS5}
\Theta[\phi]=\Theta^\prime[\phi]{{\exp[f[\phi]]-1}\over{f[\phi]}},
\end{equation}
where
\begin{equation}
\label{eq:BRS6}
f[\phi]=\sum_i\int d^4x{\delta\Theta^\prime\over{\delta\phi_i(x)}}\delta_{\rm BRS}\phi_i(x)
\end{equation}
FFBRS transformations of the type (\ref{eq:BRS3}) were used
to connect the FPEA in Lorentz-type gauges with  gauge parameter $\lambda$
to (i) the most general BRS/anti-BRS symmetric action in linear gauges,
(ii)FPEA in quadratic gauges, (iii) the FPEA in Lorentz-type gauges with another gauge parameter
$\lambda^\prime$ in \cite{jm}. It was also used to connect the former to FPEA
in axial-type gauges in \cite{jb}. We shall now summarize the results of \cite{jb}
in II C. 

\subsection{ FFBRS Transformation for Lorentz to Axial Gauge $S_{\rm eff}$}

We give results for the FFBRS transformation that connects 
the Lorentz-type gauges (See \cite{cl}) with  gauge parameter $\lambda$ to 
axial gauges (See \cite{al})) with same gauge parameter $\lambda$.
[The same calculation can be used to connect it to axial gauges 
with another  gauge
parameter $\lambda^\prime$: one simply rescales $\eta$ suitably.] 
They are obtained by integrating:
\begin{equation}
\label{eq:BRS7}
{d\phi_i(\kappa)\over{d\kappa}}=\delta_{\rm BRS}[\phi]\Theta^\prime[\phi],
\end{equation}
with
\begin{equation}
\label{eq:Thetaprimedef}
\Theta^\prime=i\int d^4x{\bar c}^\alpha
(\partial\cdot A^\alpha-\eta\cdot A^\alpha).
\end{equation}
the consequent $\Theta[\phi]$ is given by (\ref{eq:BRS5}) with
\begin{equation}
\label{eq:fdef}
f[\phi]=i\int d^4x\Biggl[{\partial\cdot A^\alpha\over\lambda}
(\partial\cdot A^\alpha-\eta\cdot A^\alpha)
+{\bar c}(\partial\cdot{\rm D}-\eta\cdot{\rm D})c^\alpha\Biggr].
\end{equation}

The meaning of these field transformations is as follows.
Suppose we begin with the vacuum expectation  value of  a gauge invariant functional
$G[\phi]$ in the Lorentz-type gauges:
\begin{equation}
\label{eq:vev}
\langle\langle G[\phi]\rangle\rangle=\int{\cal D} 
\phi G[\phi]e^{iS^L_{\rm eff}[\phi]}.
\end{equation}
Now, we perform the transformation $\phi\rightarrow\phi^\prime$ given by (\ref{eq:BRS3}).
Then we have [with $G[\phi^\prime]=G[\phi]$ by gauge invariance]
\begin{equation}
\label{eq:vevFFBRS}
\langle\langle G[\phi]
\rangle\rangle\equiv\langle\langle G[\phi^\prime]\rangle\rangle
=\int {\cal D}\phi^\prime J[\phi^\prime]G[\phi^\prime]e^{iS_{\rm eff}^L[\phi^\prime]}
\end{equation}
on account of the BRS invariance of $S_{\rm eff}^L$. 
Here $J[\phi^\prime]$ is the Jacobian
\begin{equation}
\label{eq:Jdef}
{\cal D}\phi={\cal D}\phi^\prime J[\phi^\prime].
\end{equation}
As was shown in \cite{jm}, for the special case $G[\phi]\equiv{\bf 1}$,  the
Jacobian $J[\phi^\prime]$ in (\ref{eq:Jdef}), can be replaced by 
$e^{iS[\phi^\prime]}$ where
\begin{equation}
\label{eq:S_eff^Adef}
S^L_{\rm eff}[\phi^\prime]+S_1[\phi^\prime]=S^A_{\rm eff}[\phi^\prime].
\end{equation}
As shown in Section 3, this
replacement is valid for  any gauge invariant $G[\phi]$ 
functional of  $A$. If one were to live  with vacuum expection values of 
gauge invariant observables,  the FFBRS in  \cite{jm} would be
sufficient. But as  seen in Section 3,  general Green's functions need
a modified treatment.

\section{Relation between Green's Functions for Axial-type and
Lorentz-type Gauges}

In \cite{jm},  we established a general procedure for  writing down an
FFBRS that tranforms the $W$ in one kind of a gauge choice to $W$ in
another  kind of  a gauge choice. This procedure was applied to the 
concrete example of  the construction of  an FFBRS connecting the axial-type
gauges and Lorentz-type gagues in \cite{jb}. In order to bring out
the need for  a further treatment, we first elaborate on the meaning  of  this
statement  in some detail: We note that
\begin{equation} 
\label{eq:WL}
W^L\equiv \int {\cal D}\phi e^{iS^L_{\rm eff}[\phi]}
\end{equation}
of the Lorentz-type gauges is  formally carried  over (without altering its
``value") to
\begin{equation}
\label{eq:WA}
W^A\equiv\int {\cal D}\phi^\prime e^{iS^A_{\rm eff}[\phi^\prime]}
=W^L
\end{equation}
by the FFBRS transformation
\begin{equation}
\label{eq:BRS}
\phi^\prime(x)=\phi(x)+\delta_{\rm BRS}[\phi]\Theta[\phi]
\end{equation}
constructed explicitly  in  (\ref{eq:BRS7}) - (\ref{eq:fdef}). We now
want to  use this transformation to understand how the Green's  functions
in the two gauges, and not just the vacuum-to-vacuum  ampltiudees, 
are related to each other. This
may at first  sight seem trivial. We  may expect a relation of the
kind (condensed notation used):
\begin{eqnarray}
\label{eq:naive}
& & G^A_{i_1....i_n}\equiv\int {\cal  D}\phi^\prime\prod_{r=1}^n\phi_{i_r}^\prime
e^{iS^A_{\rm eff}}[\phi^\prime]\nonumber\\
& & \stackrel{?}{=}\int {\cal D}\phi\prod_{r=1}^n\biggl(\phi_{i_r}+
\delta_{i_r;\rm BRS}[\phi]\Theta[\phi]\biggr)e^{iS^L_{\rm eff}[\phi]}\nonumber\\
& & \equiv G^L_{i_1....i_n}+\Delta G^L_{i_1....i_n},
\end{eqnarray}
where $\Delta G^L_{i_1...i_n}$, containing the terms on the right-hand
involving $\Theta$'s, gives the difference  between the $n$-point  Green's
functions $G_{1_1....i_n}$ in the two sets of gauges.
This $\Delta G^L_{i_1...i_n}$ then would be expressed in terms
of the Green's functions of the Lorentz-type gauges and may involve
additional vertices corresponding to insertions of operators 
$\delta_{\rm BRS}[\phi]$. This however, turns out to be incorrect and
the technical  reason for this is explained below.

In \cite{jm,jb}, we showed that the Jacobian for the FFBRS 
(\ref{eq:BRS}), could be replaced by a factor $\exp(iS_1)$ within the
expression for $W$ if the condition
\begin{equation}
\label{eq:condition}
\int {\cal D}\phi(\kappa)\biggl({1\over J}{dJ\over{d\kappa}}
-i{{dS_1[\phi(\kappa),\kappa]}\over{d\kappa}}\biggr)
\exp[i(S^L_{\rm eff}+S_1)]=0
\end{equation}
was fulfilled. This replacement then became valid for  $W$ (i.e. without
additional operators in the integrand of the path integral).  A priori,
it is not  obvious that if (\ref{eq:condition}) holds,
an equation of the type 
\begin{equation}
\label{eq:condition1}
\int {\cal D}\phi(\kappa){\cal O}[\phi({\kappa})]
\biggl({1\over J}{dJ\over{d\kappa}}
-i{{dS_1[\phi(\kappa),\kappa]}\over{d\kappa}}\biggr)
\exp[i(S^L_{\rm eff}+S_1)]=0,
\end{equation}
modified to include an  operator ${\cal O}[\phi(\kappa)]$
would also hold. That it  does not,  in fact, hold generally
arises from  the following  fact. The verification of (\ref{eq:condition})
in \cite{jm,jb}, made use of  the antighost equation of motion (See e.g.
the discussion below equation (3.20) of \cite{jb}).  Thus, it is  clear that
if ${\cal O}[\phi(\kappa)]$ contains ${\bar c}$, then the
procedure would fail as it involves integration by parts. In  fact, 
the procedure does not work for   any finite interval of $\kappa$ 
for any operator ${\cal O}$. This is so since ${\cal O}$ is evolving
in form (as $\kappa$ is varied) on acount of  the IFBRS transformation
of (\ref{eq:BRS7}) which  will always induce ${\bar c}$-dependence as
$\kappa$ is  varied even if  at $\kappa=0$, one  started out with
${\cal O}[A,c]$, an operator independent of  ${\bar c}$ (barring the
gauge-invariant case as discussed in  Section 2). For this reason,
the construction of relation between Green's functions for
the two type of gauges requires as elaborate a treatment as the
original  FFBRS construction itself.
We begin with a general Green's function in one of the gauges, say, the axial
gauge:
\begin{equation}
\label{eq:GA}
G=\int{\cal D}\phi^\prime{\cal O}[\phi^\prime]e^{iS^A_{\rm eff}[\phi^\prime]}.
\end{equation}
Here, the form of  ${\cal O}$ is unrestricted; so that (\ref{eq:GA})
covers arbitrary operator Green's functions as well  as arbitrary ordinary
Green's functions. For  example,  with:
\begin{equation}
\label{eq:ex1}
{\cal O}_1={A^\alpha_\mu}^\prime {A^\beta_\nu}^\prime
\end{equation}
one has the gauge boson propagator; whereas with
\begin{equation}
\label{eq:ex2}
{\cal O}_2={A^\alpha_\mu}\ ^\prime(x) 
{c^\beta}\ ^\prime(y){{\bar c}^\gamma}\ ^\prime(z)
\end{equation}
one has a 3-point Green's function; or with
\begin{equation}
\label{eq:ex3}
{\cal O}_3={F_{\mu\nu}^\alpha}^\prime(x){F^{\nu,\alpha}_\sigma}^\prime(x)
{A^\beta_\rho}^\prime(u)
{A^{\gamma,\sigma}}^\prime(w)
\end{equation}
one has the two-point Green's function of  
an operator insertion of the twist two local
operator $F^\alpha_{\mu\nu}F^{\nu,\alpha}_\sigma$,  etc.
We want to express $G$
entirely in terms of the Lorentz-type gauge Green's functions (and
possibly involving vertices from $\delta_{\rm BRS}[\phi]$). We, therefore
construct the quantity:
\begin{equation}
\label{eq:Gdef}
G(\kappa)\equiv\int {\cal D}\phi(\kappa){\cal O}[\phi(\kappa),\kappa]
e^{iS^L_{\rm eff}[\phi(\kappa)+iS_1[\phi(\kappa),\kappa]}
\end{equation}
and $define$ the form of ${\cal O}[\phi(\kappa),\kappa]$ such that
\begin{equation}
\label{eq:conG}
{dG\over d\kappa}=0.
\end{equation}
[This is exaclty analogous to the equation ${dW\over d\kappa}=0$
of \cite{jm} that related $W(1)\equiv W^A=W(0)\equiv W^L$ together.] Then
\begin{equation}
\label{eq:G(1)}
G(1)=\int {\cal D}\phi^\prime 
{\cal O}[\phi^\prime,1]e^{iS^A_{\rm eff}[\phi^\prime]}
\end{equation}
[$\phi^\prime\equiv\phi(1)$] with ${\cal O}[\phi^\prime,1]\equiv{\cal O}[\phi^\prime]$
gives the Green's function (\ref{eq:GA}), whereas it is alternately expressed
as 
\begin{equation}
\label{eq:G(0)}
G(0)=\int {\cal D}\phi\tilde{\cal O}[\phi]e^{iS^L_{\rm eff}[\phi]}=G(1),
\end{equation}
where $\tilde{\cal O}[\phi]\equiv{\cal O}[\phi(0),0]$. Equation 
(\ref{eq:G(0)}) gives the same quantity in terms 
of Lorentz gauge quantities. We,
thus, need to determine how  ${\cal O}[\phi(\kappa),\kappa]$ of 
(\ref{eq:Gdef}) should evolve so as to keep
$G(\kappa)$ independent of $\kappa$ [equation (\ref{eq:conG})].
To determine this, we perform the field transformation from $\phi(\kappa)$ to
$\phi(\kappa+d\kappa)$ via the IFBRS of (\ref{eq:BRS7}). We write, making
due use of the BRS invariance of $S^L_{\rm eff}$,
\begin{eqnarray}
\label{eq:infkappa}
& & G(\kappa)=\int {\cal D}\phi(\kappa+d\kappa){J(\kappa+d\kappa)\over J(\kappa)}
\biggl({\cal O}[\phi(\kappa+d\kappa),\kappa+d\kappa]\nonumber\\
& & -\delta_{\rm BRS}[\phi]\Theta^\prime
{\delta{\cal O}\over{\delta\phi_i}}d\kappa 
-{\partial{\cal O}\over{\partial\kappa}}d\kappa\biggr)\nonumber\\
& & \times 
e^{iS^L_{\rm eff}[\phi(\kappa+d\kappa)]
+iS_1[\phi(\kappa+d\kappa),\kappa+d\kappa]}
\times \biggl(1-i{dS_1\over{d\kappa}}d\kappa\biggr)\nonumber\\
& & =\int{\cal D}\phi[\kappa+d\kappa]
\biggl(1+{1\over J}{dJ\over{d\kappa}}d\kappa\biggr)
\biggl({\cal O}[\phi(\kappa+d\kappa),\kappa+d\kappa]\nonumber\\
& & -\delta_{\rm BRS}\Theta^\prime{\delta{\cal O}\over{\delta\phi_i}}d\kappa
-{\partial{\cal O}\over{\partial\kappa}}d\kappa\biggr)
\times\biggl(1-i{dS_1\over{d\kappa}}d\kappa\biggr)\nonumber\\
& & \times e^{iS^L_{\rm eff}[\phi(\kappa+d\kappa)]+
iS_1[\phi(\kappa+d\kappa),\kappa+d\kappa]}
\nonumber\\
& & \equiv G[\kappa+d\kappa]
\end{eqnarray}
iff
\begin{eqnarray}
\label{eq:con}
& & \int {\cal D}\phi(\kappa)\biggl(\biggl[{1\over J}{dJ\over{d\kappa}}
-i{dS_1\over{d\kappa}}\biggr]{\cal O}[\phi(\kappa),\kappa]
-\delta_{\rm BRS}\Theta^\prime{\delta{\cal O}\over{\delta\phi_i}}
-{\partial{\cal O}\over{\partial\kappa}}\biggr)\nonumber\\
& & \times
e^{iS^L_{\rm eff}[\phi(\kappa)]+iS_1[\phi(\kappa),\kappa]}\equiv 0.
\end{eqnarray}
[We have replaced  $\phi(\kappa+d\kappa)\rightarrow\phi(\kappa)$   in
(\ref{eq:con}) in view of the fact that the quantity on the left-hand 
side  is multiplied by $d\kappa$.] Thus the condition (incorrect one)
(\ref{eq:condition1}) is replaced by the correct condition (\ref{eq:con}).

We shall now simplify the condition (\ref{eq:con}) and show that this condition is
fulfilled if  a certain evolution equation is satisfied by  
${\cal O}[\phi(\kappa),\kappa]$. 
We then show how it can be solved. The procedure for  the solution to the
evolution equation will pertain to the introduction of another field
transformation and we shall show that this is an FFBRS too.

We shall now simplify the first term on the right hand side
of (\ref{eq:con}). To do this, we note that (\ref{eq:condition}) is fulfilled
and that we can use the explicit form of $S_1[\phi(\kappa),\kappa]$
and $\Theta^\prime$ of \cite{jb}
to simplify the combination ${1\over J}{dJ\over{d\kappa}}-i{dS_1\over{d\kappa}}$.
When this is done, this term reads:
\begin{eqnarray}
\label{eq:jbsim}
& & -i\int{\cal D}\phi{\cal O}[\phi(\kappa),\kappa]e^{iS^L_{\rm eff}+iS_1[\phi(\kappa),\kappa]}
\nonumber\\
& & \times\int d^4x\biggl({1\over\lambda}\kappa(\partial\cdot A-\eta\cdot A)
\biggl[(1-\kappa)Mc+\kappa \tilde M c\biggr](x)\Theta^\prime\nonumber\\
& & -\kappa{{(\partial\cdot A-\eta\cdot A)^2}\over\lambda}\biggr).
\end{eqnarray}
Now we note that 
\begin{eqnarray}
\label{eq:jbsim1}
& & \int {\cal D}\phi(\kappa){\cal O}[\phi(\kappa),\kappa]\int d^4x
{\cal F}[A(x)]\biggl[(1-\kappa)Mc+\kappa\tilde Mc\biggr](x)\Theta^\prime\nonumber\\
& & \times e^{iS^L_{\rm eff}+iS_1[\phi(\kappa),\kappa]}\nonumber\\
& & =\int {\cal D}\phi(\kappa){\cal O}[\phi(\kappa),\kappa]\int d^4x
{\cal F}[A(x)]i{\delta\over{\delta{\bar c(x)}}}e^{iS^L_{\rm eff}+iS_1}\Theta^\prime.
\end{eqnarray}
[${\cal F}[A(x)]\equiv{\kappa\over\lambda}(\partial\cdot A-\eta\cdot A)$]
Integrating by parts with respect to ${\bar c}$ and taking due account of the
anticomuting nature of ${\bar c}$ and possibly ${\cal O}$, the  above
expression equals,
\begin{eqnarray} 
\label{eq:intbypts}
& & \int  {\cal D}\phi(\kappa)\int d^4x\biggl(-i{\cal O}
\stackrel{\leftarrow}{{\delta\over{\delta{\bar c}}}}{\cal F}[A(x)]\Theta^\prime
\nonumber\\
& & 
-{\cal O}[\phi(\kappa),\kappa]{\cal F}[A(x)]
{\delta\Theta^\prime\over{\delta{\bar c}}}\biggr)
e^{iS^L_{\rm eff}+iS_1}
\end{eqnarray}
Now, the term
$\int d^4x{\delta\Theta^\prime\over{\delta{\bar c}(x)}}\kappa{\cal F}[A(x)]$
is precisely the factor that also arose in
fulfillment of (\ref{eq:con}) (i.e. when
${\cal O}[\phi(\kappa),\kappa]$ was absent) and such a term, in presence
of ${\cal O}$ cancels precisely with the last term in  (\ref{eq:jbsim}) just
as it did in (\ref{eq:condition}) in absence of ${\cal O}$.
Using the above information  in (\ref{eq:con}), the required
condition for $\kappa$-independence of $G$ reads:
\begin{eqnarray}
\label{eq:conmod}
& & \int {\cal D}\phi(\kappa) 
e^{iS^L_{\rm eff}+iS_1[\phi(\kappa),\kappa]}
\biggl(
{\partial{\cal O}\over{\partial\kappa}}
+\int {\rm D}_\mu c\Theta^\prime
{\delta{\cal O}\over{\delta A_\mu}}\nonumber\\
& & -{g\over 2}\int(fcc)\Theta^\prime
{\delta{\cal O}\over{\delta c}}+\int\biggl[{\partial\cdot A\over\lambda}
+\kappa{{(\eta\cdot A -\partial\cdot A)}\over\lambda}\biggr]\Theta^\prime
{\delta{\cal O}\over{\delta{\bar c}}}\biggr)=0
\end{eqnarray}
So far we have not spelled out the $\kappa$-dependence of  ${\cal O}$.
Now, if  we $construct$ an ${\cal O}[\phi(\kappa),\kappa]$ which
satisfies:
\begin{eqnarray} 
\label{eq:conO}
& & 
{\partial{\cal O}\over{\partial\kappa}}
+\int {\rm D}_\mu c\Theta^\prime
{\delta{\cal O}\over{\delta A_\mu}}\nonumber\\
& & -{g\over 2}\int(fcc)\Theta^\prime
{\delta{\cal O}\over{\delta c}}+\int\biggl[{\partial\cdot A\over\lambda}
+\kappa{{(\eta\cdot A -\partial\cdot A)}\over\lambda}\biggr]\Theta^\prime
{\delta{\cal O}\over{\delta{\bar c}}}=0,
\end{eqnarray}
then (\ref{eq:conmod}) would automatically be satisfied, 
thus leading to (\ref{eq:G(0)}). Thus, we have to know how to
solve (\ref{eq:conO}) to obtain ${\cal O}[\phi(\kappa),\kappa]$. To this
end, consider the $same$ function ${\cal O}$ with  a different
argument $\tilde\phi(\kappa)$, ${\cal O}[\tilde\phi(\kappa),\kappa]$
where $\tilde\phi(\kappa)$ is
defined via a new set of evolution equations: 
\begin{eqnarray}
\label{eq:modBRS}
& & {d\tilde A_\mu(\kappa)\over{d\kappa}}={\rm D}_\mu[\tilde A]\tilde c\
\Theta^\prime [\tilde\phi(\kappa)]\nonumber\\
& & {d\tilde c\over{d\kappa}}=-{g\over 2}f\tilde c(\kappa)\tilde c(\kappa)
\Theta^\prime[\tilde\phi(\kappa)]\nonumber\\
& & {d\tilde{\bar c}(\kappa)\over{d\kappa}}
={{\partial\cdot\tilde A(\kappa)+\kappa
(\eta\cdot\tilde A-\partial\cdot\tilde A)}\over\lambda}
\Theta^\prime[\tilde\phi(\kappa)],
\end{eqnarray}
or in short,
\begin{equation}
\label{eq:modBRS1}
{d\tilde\phi(\kappa)\over{d\kappa}}\equiv\tilde\delta
[\tilde\phi(\kappa),\kappa]
\Theta^\prime[\tilde\phi(\kappa)],
\end{equation}
together with the boundary condition:
\begin{equation}
\label{eq:boundcon}
\tilde\phi(1)=\phi^\prime=\phi(1).
\end{equation}
The the condition (\ref{eq:conO}) when expressed for 
${\cal O}[\tilde\phi(\kappa),\kappa]$ instead of 
${\cal O}[\phi(\kappa),\kappa]$ as:
\begin{equation}
\label{eq:conOtilde}
{\partial{\cal O}[\tilde\phi(\kappa),\kappa]\over{\partial\kappa}}
+\int \tilde\delta_{i\ BRS}[\tilde\phi(\kappa),\kappa]
\Theta^\prime[\tilde\phi(\kappa)]
{\delta{\cal O}[\tilde\phi(\kappa),\kappa]\over{\delta\tilde\phi_i(\kappa)}}=0,
\end{equation}
i.e.
\begin{equation}
\label{eq:conOtilde1}
{d{\cal O}[\tilde\phi(\kappa),\kappa]\over{d\kappa}}\equiv 0.
\end{equation}
Now, in view of the fact that
\begin{equation}
\label{eq:bc1}
{\cal O}[\tilde\phi(1),1]={\cal O}[\phi(1),1]={\cal O}[\phi^\prime],
\end{equation}
we find 
\begin{equation}
\label{eq:bc2}
{\cal O}[\tilde\phi(\kappa),\kappa]={\cal O}[\phi^\prime].
\end{equation}

Equation (\ref{eq:bc2}) tells us how the function 
${\cal O}[\phi(\kappa),\kappa]$ should evolve: we solve
(\ref{eq:modBRS1}) for $\phi^\prime$ in terms of $\tilde\phi(\kappa)$,
express 
${\cal O}[\phi^\prime]\equiv{\cal O}[\phi^\prime(\tilde\phi(\kappa),\kappa)]$
$={\cal O}[\tilde\phi(\kappa),\kappa]$. This gives us the function ${\cal O}$.
In this we replace the argument $\tilde\phi\rightarrow\phi$ to obtain 
${\cal O}[\phi(\kappa),\kappa]$ which then will solve (\ref{eq:conO}).
The value of ${\cal O}[\phi(\kappa),\kappa]$ at $\kappa=0$, 
i.e., ${\cal O}[\phi(0), 0]$ will then give us the function
$\tilde{\cal O}[\phi]$ of (\ref{eq:G(0)}) involved in the
expression of $G(1)$ in terms of the Lorentz gauge quantities. Thus
the evolution of ${\cal O}[\phi(\kappa),\kappa]$ with $\kappa$ is easy to
obtain if the IFBRS (\ref{eq:modBRS1}) is solved. The IFBRS
of (\ref{eq:modBRS1}) differs
from the IFBRS of (\ref{eq:BRS4}) in that the 
transformation for $\tilde{\bar c}$ involves the 
$\tilde\delta[\phi(\kappa),\kappa]$ and
is  explicitly  $\kappa$-dependent. The integration the IFBRS proceeds the 
same way as the basic IFBRS (\ref{eq:BRS4}) as done in \cite{jm}; the only 
complication  being the $\kappa$-dependent
$\tilde\delta_{\rm BRS}[\tilde\phi(\kappa),\kappa]$ involved
in ${d\bar c\over{d\kappa}}$. The integration is given in
appendix A. The result is 
\begin{eqnarray}
\label{eq:result1}
& & \phi^\prime=\phi+\biggl(\tilde\delta_1[\phi]\Theta_1[\phi]
+\delta_2[\phi]\Theta_2[\phi]\biggr)\Theta^\prime[\phi]
\nonumber\\
& & \equiv\phi+\delta\phi[\phi]
\end{eqnarray}
Using (\ref{eq:result1}), (\ref{eq:G(0)}), (\ref{eq:bc1}) and
(\ref{eq:bc2}), we obtain the following  result:
\begin{eqnarray}
\label{eq:result2}
& & 
G(1)=G(0)=\int {\cal D}\phi\tilde{\cal O}[\phi]e^{iS^L_{\rm eff}[\phi]}\nonumber\\
& & \int{\cal D}\phi{\cal O}\biggl(\phi+\delta\phi[\phi]\biggr)
e^{iS^L_{\rm eff}[\phi]}.
\end{eqnarray}
In view of the nilpotency of $\delta[\phi]$, this leads to
\begin{equation}
\label{eq:result3}
\int{\cal D}\phi{\cal O}[\phi]e^{iS^L_{\rm eff}[\phi]}+
\int{\cal D}\phi\delta\phi[\phi]
{\delta{\cal O}\over{\delta\phi_i}} e^{iS^L_{\rm eff}}.
\end{equation}
Further, as done in appendix A, the last term can be cast in a neat form; 
so that (\ref{eq:result3}) can be written as:
\begin{equation}
\label{eq:AtoL2}
\langle {\cal O}\rangle_A=\langle{\cal O}\rangle_L+\int_0^1 d\kappa\int D\phi  
\biggl(\tilde\delta_1[\phi]+\kappa\tilde\delta_2[\phi]
\biggr)\Theta^\prime[\phi]
{\delta{\cal O}\over{\delta\phi}} e^{iS^M_{\rm eff}}.
\end{equation}

Our aim in this
work was to establish formally the link between two gauges considered. This has
been done in (\ref{eq:result3}) and (\ref{eq:AtoL2}). In this
work, we shall content ourselves with some comments
on concrete calculations. Concrete
evaluation of (\ref{eq:AtoL2}) can be carried out in two  ways.
While the one based on (\ref{eq:AtoL2}) is much superior, we shall
enumerate both for formal reasons.
(I) We can look upon the integrand on the right hand side as 
an expansion in $\kappa$ (and carry out the $\kappa$ integration). 
Then each  term
gives a Green's function of the operator ${\cal O}$ (and its BRS variation) in
Lorentz-type gauges. We can further regard 
each term in the expansion as an expansion in $g$. 
Then to a given desired order only a
finite number of terms in each need be kept. To any given
order in $g$, the infinite terms have
however to be summed.  This however can be avoided with the help of
the alternate expresion (\ref{eq:AtoL2}) which turns
out to be much superior for practical purposes. 
(II) We can alternately regard the evaluation of the
functional integral on the right hand side of (\ref{eq:AtoL2}) in terms
of the vertices and the propagators
of the interpolating mixed gauge action $S^M_{\rm eff}$. 
This approach has many technical 
advantages. The last term on the right hand side of
(\ref{eq:AtoL2}) now consists of usual Feynman diagrams, with one difference:
the propagators of ghost and gauge fields are now 
$\kappa$-dependent and a final (overall) $\kappa$
integration is to be preformed. 
To any given order in $g$, there are only a finite number of
Feynman diagrams to be evaluated on the right hand side. If, for example
$O$ is a local polynomial operator, these are the
Feynman diagrams with one insertion each of two
local polynomial
operators (or integrated local density), 
$\delta_{\rm BRS}[\phi_i]{\delta O\over{\delta\phi_i}}$
and $\int d^4x {\bar c}(x)(\partial\cdot A-\eta\cdot A)(x)$, and
can be evaluated by usual techniques. If, on the other hand, $O$
is a product of $n$ elementary fields at distinct space time points (such as
in (\ref{eq:ex1}) and (\ref{eq:ex2})), then the right hand side has (a finite number of)
Feynman diagrams corresponding to the $(n-1)$-point functions with
one insertion each of $\delta_{\rm BRS}[\phi_i]$ and
$\int d^4x {\bar c}(\partial\cdot A-\eta\cdot A)(x)$.
[We shall give a simple
 example of this calculation in Section 4] 
We can use such an expansion (especially the approach II) 
to correlate the axial gauge propagator in terms of Lorentz gauge
quantities. Knowing how to deal with
the Lorentz gauge calculations should throw
direct light on how to deal with axial gauge calculations especially
the prescription for the ${1\over{\eta\cdot k}}$-type singularities in
axial propagator. It should also help in 
resolving number of  existing problems with light cone gauge calculations. This
work is in progress \cite{BRS2,progress}.

We expect such relations to  resolve the discrepancy reported between
the anomalous dimensions of physical obervables in the two
sets of gauges \cite{progress}. We leave the issue to a further  publication.

\section{An Example}

In this section, we shall give a simple example of the relation 
(\ref{eq:AtoL2}).  Consider for example,
\begin{equation}
\label{eq:exreln1}
O[\phi]=A^\alpha_\mu(x)A^\beta_\nu(y).
\end{equation}
Then
\begin{equation}
\label{eq:exreln2}
\langle O[\phi]\rangle_A\equiv
\langle A^\alpha_\mu(x) A^\beta_\nu(y)\rangle=iG^{A\ \alpha\beta}_{\mu\nu}(x-y)
\end{equation}
is (for the connected part) the axial gauge propagators. In
obvious notations
\begin{eqnarray}
\label{eq:exres}
& & iG^{A\ \alpha\beta}_{\mu\nu}(x-y)=iG^{L\ \alpha\beta}_{\mu\nu}(x-y)
+i\int_0^1 d\kappa\int{\cal D}\phi e^{iS_{\rm eff}^M[\phi,\kappa]-i
\epsilon\int(A^2/2-{\bar c}c)d^4x}\nonumber\\
& & \times
\biggl(({\rm D}_\mu c)^\alpha(x)A^\beta_\nu(y)
+A^\alpha_\mu(x)
({\rm D}_\nu c)^\beta(y)\biggr)
\int d^4z{\bar c}^\gamma(z)(\partial\cdot A^\gamma-\eta\cdot A^\gamma)(z)
\end{eqnarray}
The right hand side consists of one point functions
of one insertion each of two
local operators (or integrated local density) ${\rm D}_\mu c$ and
$\int d^4x{\bar c}(\partial\cdot A-\eta\cdot A)d^4x$. To any
finite order, such terms can be evaluated by drawing the appropriate
Feynman diagram whose propagators and vertices arise from
$S^M_{\rm eff}[\phi,\kappa]$. The propagators are
now $\kappa$-dependent. We expect the results
(\ref{eq:result3}) and (\ref{eq:AtoL2}) to be useful in this manner to be
able to solve the number of problems mentioned in the
Introduction (last-but-one paragraph). We now make brief comments on one such
application as an example.

Eqauation (\ref{eq:exres}) leads to, for zero loop case,
\begin{eqnarray}
\label{eq:exres1}
&  & G^{0 A\ \alpha\beta}_{\mu\nu}(x-y)=G^{0 L\ \alpha\beta}_{\mu\nu}(x-y)
\nonumber\\
& & -i\int_0^1 d\kappa\biggl[-i\partial_\mu^x
\tilde G^{0M}(x-y)(\partial_z^\sigma-\eta^\sigma)
\tilde G^{0M\ \alpha\beta}_{\sigma\nu}
+(\mu,x,\alpha)\leftrightarrow(\nu,y,\beta)\biggr].
\nonumber\\   
& &
\end{eqnarray}  
The last term on  the right hand side  involves $\kappa$-dependent
functions for ghost and gauge fields:
\begin{equation}
\label{eq:eps5}
\tilde G^{0 M}(x-y)
=\int d^4q{e^{-iq\cdot (x-y)}\over{(\kappa-1)q^2-i\kappa
q\cdot\eta-i\epsilon}}
\end{equation}
and   
\begin{equation}
\label{eq:eps6}
\tilde G^{0 M\ \alpha\beta}_{\sigma\nu}(x-y)
=\delta^{\alpha\beta}\int d^4k e^{-ik\cdot (x-y)} 
\tilde G^{0M}_{\sigma\nu}(k)
\end{equation}
with
\begin{eqnarray}   
\label{eq:pres6}
& & \tilde G^{0 M}_{\mu\rho}(k)
=-{1\over{k^2+i\epsilon}}\Biggl[g_{\mu\rho}+\nonumber\\
& &
{{\biggl(\biggl[[(1-\kappa)^2-\lambda]-{\eta^2\kappa^2\over{k^2+i\epsilon}}
\biggr]k_\mu k_\rho
-i\kappa(1-\kappa)k_{[\mu}\eta_{\rho]}
+{\kappa^2\eta\cdot k\over{k^2+i\epsilon}}k_{[\mu}\eta_{\rho]_+}
+i{\kappa^2\epsilon\over{k^2+i\epsilon}}\eta_\mu\eta_\rho\biggr)}
\over{\biggl(-{\kappa^2\over{(k^2+i\epsilon)}}
\biggl[(\eta\cdot k)^2-\eta^2k^2+(k^2+\eta^2)(k^2+i\epsilon)\biggr]
+2k^2\kappa
-i\epsilon\lambda-k^2\biggr)}}\Biggr].
\nonumber\\
& & 
\end{eqnarray}
[It should be emphasized that (\ref{eq:eps5}), (\ref{eq:eps6}) are only
intermediate objects occurring in calculations
and are $not$ the actual   ghost   and gauge
propagators (even in intermediate gauges)
as the latter must be evaluated ultimately with
a term like $\epsilon O^\prime_1[\phi^\prime,\kappa]$ in the exponent.] We
obtain:
\begin{eqnarray}
\label{eq:Result1}
& & \tilde G^{0 A}_{\mu\nu}-\tilde G^{0 L}_{\mu\nu}
={-i\over{(k^2+i\epsilon)^2(1-i\xi_1-i\xi_2)(1-i\xi_2+\xi_1^2+i\xi_2\xi_3)}}
\nonumber\\
& & \times\int_0^1 d\kappa{\Biggl[k_\mu k_\nu\biggl(\kappa
+\biggl[{i\lambda-\xi_1(1-\lambda)\over{\xi_1+i\xi_3}}\biggr]\biggr)
(\xi_1+i\xi_3)+\eta_\mu k_\nu
\biggl(\kappa+\biggl[{1-i\xi_2(1-\lambda)\over{-1-i\xi_1+i\xi_2}}\biggr]\biggr)
(-1-i\xi_1+i\xi_2)\Biggr]
\over{(\kappa-a_1)(\kappa^2-2\gamma\kappa+\beta)}}\nonumber\\
& & +(k\rightarrow-k,\ \mu\leftrightarrow\nu)
\end{eqnarray}
with                                                                     
\begin{eqnarray}
\label{eq:defins}
& & \xi_1\equiv{\eta\cdot k\over{k^2+i\epsilon}};
\ \xi_2\equiv{\epsilon\over{k^2+i\epsilon}};
\ \xi_3\equiv{\eta^2\over{k^2+i\epsilon}};\nonumber\\
& & a_1\equiv{1\over{1-i\xi_1-i\xi_2}};\nonumber\\
& &
\gamma\equiv{(1-i\xi_2)\over{1-i\xi_2+\xi_1^2+i\xi_2\xi_3}}
\equiv{{1-i\xi_2}\over D}\nonumber\\
& & \beta\equiv{1+i\xi_2(\lambda-1)\over{1-i\xi_2+\xi_1^2+i\xi_2\xi_3}}
=\gamma +{i\xi_2\lambda\over D}.
\end{eqnarray}
For $|\eta\cdot k|>>\epsilon$
one can show that (\ref{eq:Result1}) leads
to the usual behavior of the axial propagator (See \cite{BRS2,progress}), 
which then reads:
\begin{equation}
\label{eq:eps=0}
\tilde G^{0 A}_{\mu\nu}-\tilde G^{0 L}_{\mu\nu}=
-{1\over k^2}k_\mu k_\nu
\biggl({(\lambda k^2+\eta^2)\over{(\eta\cdot k)^2}}+{(1-\lambda)\over k^2}
\biggr) + {k_{[\mu}\eta_{\nu]_+}\over{k^2\eta\cdot k}}.
\end{equation}
Equation (\ref{eq:Result1}) has been used to deal with the 
singularity structure near $\eta\cdot k=0$
\cite{BRS2,progress}.

\section{Conclusions and Further Directions}

In this work, we addressed the
problem of relating calculations in two sets of uncorrelated
gauges. We took for concreteness the axial and the Lorentz-type
gauges from the point of view of their common usage. We used the results
of \cite{jm} applied to the concete case of FFBRS for axial  and Lorentz-
type gauges obtained in \cite{jb}. We established a procedure for
relating arbitrary Green's functions in the 
two sets of gauges. We showed that this involved another but
related  FFBRS, obtained
by intregration of an IFBRS as in \cite{jm}.
We found that the final result could be put in a neat form
(\ref{eq:result3}) or (\ref{eq:AtoL2}). Form
(\ref{eq:result3}) is particularly
useful from calculational point of vew.
We expect our results to be useful in (i) deriving the correct prespription for 
${1\over{\eta\cdot k}}$-type singularities
in axial gauges; (ii) providing insights into problems associated
with existing prescriptions in axial/light cone gauges;
(iii) resolving existing discrepancies in the two sets of gauges.

\section*{Acknowledgements}

One of use [SDJ] would like to thank for its hospitality the
Erwin Schrodinger Institute of Mathematical Physics,
Vienna, where part of the work was done.

\appendix
\section{Modified FFBRS}
\setcounter{equation}{0}
\seceqaa

For  the modified IFBRS,
we wish to show that it can be integrated along
the lines of \cite{jm} (Section 3). As done there,
we can write with modification in $f$ of (3.6) of \cite{jm}:
\begin{equation}
\label{eq:f12def1}
f[\tilde\phi,\kappa]\equiv 
f_1[\tilde\phi]+\kappa f_2[\tilde\phi].
\end{equation}
Then,
\begin{equation}
\label{eq:derTheta'1}
{{d\Theta^\prime[\tilde\phi(\kappa)]}\over{d\kappa}}=
(f_1[\tilde\phi]+\kappa f_2[\tilde\phi])
\Theta^\prime[\tilde\phi(\kappa)].
\end{equation}
Following \cite{jm}, we note $f_i[\tilde\phi(\kappa)]
\Theta^\prime[\tilde\phi(\kappa)]
=f_i[\tilde\phi(0)] \Theta^\prime[\tilde\phi(\kappa)]\equiv f_i [\phi]
\Theta^\prime[\tilde\phi(\kappa)](i=1,2)$, one gets:
\begin{equation}
\label{eq:derTheta'2}
{{d\Theta^\prime[\tilde\phi(\kappa)]}\over{d\kappa}}=
(f_1[\phi]+\kappa f_2[\phi])
\Theta^\prime[\tilde\phi(\kappa)].
\end{equation}
Integrating  (\ref{eq:derTheta'2}) from $\kappa=0$ to $\kappa=\kappa$,
\begin{eqnarray}
\label{eq:int1}
& & 
\Theta^\prime[\tilde\phi(\kappa)]=\Theta[\phi]\exp\biggl(\int_0^\kappa 
f[\phi^\prime(\kappa)]d\kappa^\prime\biggr)\nonumber\\
& & 
=\Theta[\phi]\exp\biggl(\kappa f_1[\phi]
+{\kappa^2\over 2} f_2[\phi]\biggr).
\end{eqnarray}
Similarly, one writes
\begin{eqnarray}
\label{eq:derphitilde}
& & {{d\tilde\phi(\kappa)}\over{d\kappa}}=
(\tilde\delta_1[\tilde\phi(\kappa)]
+\kappa\tilde\delta_2[\tilde\phi(\kappa)])
\Theta^\prime[\tilde\phi(\kappa)]\nonumber\\
& & =
(\tilde\delta_1[\phi]
+\kappa\tilde\delta_2[\phi])
\Theta^\prime[\tilde\phi(\kappa)].
\end{eqnarray}
Integrating (\ref{eq:derphitilde}) from $\kappa=0$ to $\kappa=1$, one gets:
\begin{equation}
\label{eq:int2}
\phi^\prime=\phi+(\tilde\delta_1[\phi]\Theta_1[\phi]
+\tilde\delta_2\Theta_2[\phi])\Theta^\prime[\phi],
\end{equation}
where 
\begin{equation}
\label{eq:Theta12def}
\Theta_{1,2}[\phi]\equiv\int_0^1 d\kappa (1,\kappa)\exp\biggl(\kappa f_1[\phi]
+{\kappa^2\over 2}f_2[\phi]\biggr).
\end{equation}
For the modified FFBRS of Section 3, 
\begin{eqnarray} 
\label{eq:f12def2}
& & f_1[\phi]\equiv i\int d^4x\biggl[{\partial\cdot A^\alpha\over\lambda}
(\partial\cdot A^\alpha-\eta\cdot A^\alpha)+{\bar c}
(\partial\cdot{\rm D}-\eta\cdot{\rm D})c\biggr]
\nonumber\\
& & f_2[\phi]\equiv -{i\over\lambda}
\int d^4x (\partial\cdot A^\alpha-\eta\cdot A^\alpha)^2
\end{eqnarray}

Now we apply the FFBRS of (\ref{eq:int2}) to the problem at hand.
Consider the vev of ${\cal O}$ in the axial gauge:
\begin{equation}
\label{eq:vevaxial}
\int {\cal D}\phi^\prime {\cal O}[\phi^\prime]e^{iS^A_{\rm eff}}.
\end{equation}
Now,
\begin{equation}
\label{eq:AtoL}
{\cal O}[\phi^\prime]\equiv{\cal O}[\phi
+(\tilde\delta_1\Theta_1 + \tilde\delta_2\Theta_2)\Theta^\prime]
={\cal O}[\phi]
+(\tilde\delta_1\Theta_1 + \tilde\delta_2\Theta_2)\Theta^\prime
{\delta{\cal O}\over{\delta\phi}}.
\end{equation}
We substitute (\ref{eq:AtoL}) in (\ref{eq:vevaxial}) to obtain
\begin{eqnarray}
\label{eq:AtoL1}
& & G^A_O\equiv
\int{\cal D}\phi^\prime {\cal  O}[\phi^\prime] e^{iS^A_{\rm eff}}\nonumber\\
& & \int {\cal D}\phi {\cal O}[\phi]e^{iS^L_{\rm eff}}+\int{\cal D}\phi
(\tilde\delta_1\Theta_1 + \tilde\delta_2\Theta_2)\Theta^\prime
{\delta{\cal O}\over{\delta\phi}} e^{iS^L_{\rm eff}}.
\end{eqnarray}
We now note the forms of $\Theta_1$ and $\Theta_2$ in (\ref{eq:Theta12def})
and that
\begin{equation}
\label{eq:Mgauge}
iS^L_{\rm eff}+\kappa f_1[\phi]
+{\kappa^2\over 2}f_2[\phi]\equiv iS^M_{\rm eff}.
\end{equation}
This leads us to
\begin{equation}
\label{eq:AtoL2repeat}
\langle {\cal O}\rangle_A=\langle{\cal O}\rangle_L+\int_0^1 d\kappa\int D\phi  
\biggl(\tilde\delta_1[\phi]+\kappa\tilde\delta_2[\phi]
\biggr)\Theta^\prime[\phi]
{\delta{\cal O}\over{\delta\phi}} e^{iS^M_{\rm eff}}.
\end{equation}


\begin{thebibliography}{99}
\bibitem{cl} Ta-Pei Cheng and Ling Li,
{\it Gauge Theory of Elementary Particle Physics} (Clarendon, Oxford 1984)
\bibitem{lm} Leibrandt G, Phys Rev D ${\bf 29}$, 1699 (1984)\\
Mandelstam S, Nucl Phys B ${\bf 213}$, 149 (1983)
\bibitem{l} See for example references in G. Leibrandt, Rev Mod Phys
${\bf 59}$, 1067 (1987)
\bibitem{s} M.A.Shifman, Nucl Phys B${\bf 173}$,
13 (1980)
\bibitem{dn} N.G.Deshpande and N.Narerimonford Phys Rev D 
${\bf 27}$, 1165 (1983)
\bibitem{al} E. Abers and B.W.Lee, Phys. Rep. C ${\bf 9}$, 1 (1973), and
references therein
\bibitem{h} For some formal results, see M.Henneaux, Phys. Rep ${\bf 126}$, 
1 (1985), and references therein
\bibitem{jb1} See references quoted in ref. 10 below
\bibitem{jm} S.D.Joglekar and B.P.Mandal, Phys Rev D 
${\bf 51}$, 1919 (1995)
\bibitem{jb} R.S.Bandhu and S.D.Joglekar J.Phys.A-Math and General ${\bf 31}$,
4217 (1998)
\bibitem{talk} S.D.Joglekar in {\it Finite Field-dependent
BRS (FFBRS) Transformations and Axial Gauges}; invited talk at the
conference titled {\it Theoretical Physics Today: Trends and
Perspectives} held at IAS Calcutta, April 1998, published in
Ind. J. Physics 73B(2), 137-145 (1999)
\bibitem{BRS2} S.D.Joglekar and A.Misra, hep-th/9904107, to appear in Mod. Phys. Lett. A 
\bibitem{progress} S.D.Joglekar and A.Misra, hep-th/9909123 
\end{thebibliography}
\end{document}